\documentclass[iop]{emulateapj} 
\usepackage{natbib}
\usepackage{epstopdf}
\usepackage[]{graphicx}

\usepackage[T1]{fontenc}
\usepackage{aecompl}
\newcommand{\hiir}{H~{\scshape ii}~region}
\newcommand{\hiirs}{H~{\scshape ii}~regions}
\newcommand{\msun}{M$_{\sun}$}
\newcommand{\um}{$\mu$m}
\newcommand{\degree}{^{\circ}}
\newcommand{\reff}{R$_{\rm eff}$}
\newcommand{\nbubs}{3599}
\newcommand{\nclumps}{10285}

\begin{document}


\title{The Milky Way Project and ATLASGAL: The distribution and physical properties of cold clumps near infrared bubbles}


\author{Sarah Kendrew}
\affil{University of Oxford, Department of Astrophysics, Denys Wilkinson Building, Keble Road, Oxford OX1 3RH, United Kingdom}
\affil{Max-Planck-Institut f\"{u}r Astronomie, K\"{o}nigstuhl 17, 69117 Heidelberg, Germany}
\email{sarah.kendrew@physics.ox.ac.uk}

\author{Henrik Beuther}
\affil{Max-Planck-Institut f\"{u}r Astronomie, K\"{o}nigstuhl 17, 69117 Heidelberg, Germany}

\author{Robert Simpson}
\affil{Google UK, Belgrave House, 76 Buckingham Palace Road, London SW1W 9TQ, United Kingdom}
\affil{University of Oxford, Department of Astrophysics, Denys Wilkinson Building, Keble Road, Oxford OX1 3RH, United Kingdom}

\author{Timea Csengeri}
\affil{Max-Planck-Institut f\"{u}r Radioastronomie, Auf dem H\"{u}gel 69, 53121 Bonn, Germany}

\author{Marion Wienen}
\affil{Max-Planck-Institut f\"{u}r Radioastronomie, Auf dem H\"{u}gel 69, 53121 Bonn, Germany}

\author{Chris.~J. Lintott}
\affil{University of Oxford, Department of Astrophysics, Denys Wilkinson Building, Keble Road, Oxford OX1 3RH, United Kingdom}

\author{Matthew S. Povich}
\affil{Cal Poly Pomona, Department of Physics and Astronomy, 3801 West Temple Ave, Pomona, CA 91768, USA}

\author{Chris Beaumont}
\affil{Harvard-Smithsonian Center for Astrophysics, 60 Garden Street, Cambridge, MA 02138, USA}

\author{Fr\'{e}d\'{e}ric Schuller}
\affil{ESO, Alonso de C\'{o}rdova 3107, Vitacura, Casilla 19001, Santiago de Chile, Chile}




\begin{abstract}
We present a statistical study of the distribution and physical properties of cold dense material in and around the inner Galactic Plane near infrared bubbles as catalogued by the Milky Way Project citizen scientists. Using data from the ATLASGAL 870~\micron~survey, we show that 48 $\pm$ 2\% of all cold clumps in the studied survey region ($|l| \le 65\degree$, $|b| \le 1\degree$) are found in close proximity to a bubble, and 25 $\pm$ 2\% appear directly projected towards a bubble rim. A two-point correlation analysis confirms the strong correlation of massive cold clumps with expanding bubbles. It shows an overdensity of clumps along bubble rims that grows with increasing bubble size, which shows how interstellar medium material is reordered on large scales by bubble expansion around regions of massive star formation. The highest column density clumps appear resistent to the expansion, remaining overdense towards the bubbles' interior rather than being swept up by the expanding edge. Spectroscopic observations in ammonia show that cold dust clumps near bubbles appear to be denser, hotter and more turbulent than those in the field, offering circumstantial evidence that bubble-associated clumps are more likely to be forming stars. These observed differences in physical conditions persist for beyond the region of the bubble rims.
\end{abstract}


\keywords{Infrared: ISM; ISM: bubbles, HII regions; Stars: formation, massive; Submillimeter}



\section{Introduction}\label{sec:intro}

Feedback from massive stars has far-reaching effects on a galaxy's interstellar medium (ISM) and is thought to play a crucial role in the life cycle of molecular material in the Galaxy and the regulation of star formation on global scales~\citep{Hopkins2011, Hopkins2014, Dale2011}. Ionizing radiation and stellar winds from massive stars and clusters carve out low-density bubbles and shells, readily visible at infrared (IR) wavelengths due to emission from polycyclic aromatic hydrocarbons (PAHs) or heated dust~\citep{Churchwell2006, Churchwell2007}. These bubbles are thus ideal laboratories for studying the effects of massive stellar feedback on the surrounding material, and infrared observations of these objects form a good complement to observations at other wavelengths, e.g. in optical or radio.

\subsection{Feedback from massive stars and clusters}

Feedback phenomena surrounding newly formed massive stars and clusters have received much interest in the literature in recent years, particularly in the context of their effect on further star formation activity within the natal cloud and on global galactic scales~\citep{Krumholz2009, Krumholz2014, Dale2011, Faucher2013, Tremblin2014a}. The overall effect is determined by a complex interplay of forces acting on a range of spatial and temporal scales: ionising radiation, radiation pressure and stellar wind momentum from the young stars interact with the cloud material of a given density, size, self-gravity and magnitude of turbulence. 

Numerous studies report observational evidence of triggered star formation near massive young stars or clusters, whereby the energy injected into the ISM causes the compression, fragmentation and gravitational collapse of cold dense gas, or of pre-existing condensations within the natal cloud, and ultimately the formation of a new generation of young stars~\citep[e.g.][]{deharveng05,deharveng10,martins10, Zavagno2010a}. Other proposed causes of triggering include supernova explosions~\citep{Phillips2009}; massive stellar winds~\citep{Castor1975}; protostellar outflows~\citep{Barsony1998}; and spiral density waves~\citep{Roberts1969} and galaxy-galaxy tidal interactions~\citep{Woods2006} on Galactic scales. Establishing whether the 2nd generation star formation was actively triggered by feedback or simply taken place \emph{in situ}, is however very challenging and requires a comprehensive multi-wavelength approach~\citep{Dale2015}. 

In addition, massive stellar feedback may \emph{disrupt} further star formation locally, removing gas from the protostars' surroundings, halting collapse or accretion, or disrupting the parent cluster~\citep{Boily2003, Dale2005, Hopkins2011, Bastian2006}. Turbulent energy injected into the ISM by ionizing radiation and winds can prevent further protostellar collapse over larger distances~\citep{Matzner2002}. The role of turbulence was studied in detail by~\citet{McKee2007}; recently~\citet{Kruijssen2014} identify it as a crucial regulator of a molecular cloud's star formation efficiency. The importance of bubbles was highlighted in recent work by~\citet{Rahman2010, Lee2012}, who find that the feedback energy from expanding bubbles in star forming complexes is a major driver of turbulence in the ISM of the inner Milky Way; possibly more important than that from supernovae. Both triggering and quenching of star formation by massive stellar feedback are observed in smoothed particle hydrodynamics (SPH) simulations by~\citet{Dale2005, Dale2007a} and~\citet{Walch2012}.

\subsection{Statistical studies of massive stellar feedback}

The above observational studies and simulations have shown that identifying young stars formed through triggering is very challenging for any given star forming region. Such studies in addition do not address the question of the Galactic-scale importance of feedback on star formation. One solution to this problem is to correlate star forming populations statistically.~\citet{thompson12} demonstrated the use of correlation functions for studying the importance of triggered star formation on Galactic scales, using the~\citet{Churchwell2006, Churchwell2007} catalogs of infrared bubbles, catalogued from GLIMPSE survey images~\citep{Benjamin2003a}, and the Red MSX Source Survey (RMS) catalog of young stellar objects (YSOs) and compact \hiirs~; they find a clear overdensity of RMS sources along the rims of bubbles.

In~\citet{Kendrew2012} we extended this analysis to include the order-of-magnitude larger sample of IR bubbles catalogued by the Milky Way Project (MWP) citizen science website. Our results were broadly consistent with those of~\citet{thompson12}, 
showing that 22 $\pm$ 2\% of massive young stars are located near the rims of bubbles, with the angular correlation function showing a statistically significant overdensity of massive YSOs (MYSOs) near the rims of the largest bubbles.These findings suggest an \emph{upper limit} on the fraction of massive stars that may form as a direct result of stellar feedback. The datasets used in this analysis come with two important limitations. 

First, the MSX beam size of 18\arcsec~is similar to the median bubble size ($\sim$36\arcsec), and both surveys are conducted at similar wavelengths, thus probing similar physics. It is therefore clear that a non-negligible subset of catalogue sources are found in both sets. A strong correlation may therefore well mean the sources are the same, rather than associated. Second, as the RMS and GLIMPSE surveys overlap in wavelengths, the data trace sources over similar evolutionary stages, and a similarly wide range. We therefore cannot establish a reliable evolutionary sequence based on the observed associations between these datasets.

In addition, interpretation of the observed trend with bubble sizes is challenging without knowledge of bubble distances. For a sample of 185 bubbles with counterparts in the~\citet{Anderson2009} catalogue of~\hiirs~with known distances, no distance-physical size correlation is identified (see Fig. 19 in~\citet{simpson_dr1}), suggesting that the observed overdensity of MYSOs near the rims of the largest bubbles is not simply a distance or completeness effect. 

In this paper we address one of these limitations by correlating the MWP bubbles catalogue with submillimeter wavelength data, specifically the~\citet[][\citet{Csengeri2014} hereafter]{Csengeri2014} catalog of cold dense clumps from the APEX Telescope Large Area Survey of the Galaxy~\citep[ATLASGAL]{Schuller2009}, and a spectroscopic follow-up study by~\citet{Wienen2012} in NH$_3$ used to derive kinematic distance estimates. The ATLASGAL source catalogue of \citet{Csengeri2014} consists of almost 11000 cold dust clumps detected in 870~\micron~continuum emission, representing regions likely to host predominantly massive star formation, at earlier evolutionary stages than the RMS sample. Use of this dataset also broadens the study of Galactic-scale feedback by focusing not only on those regions that have already started forming stars, but on the dense gas of the ISM on a global scale, quiescent as well as star-forming.

\section{Data catalogs}\label{sec:data}

\subsection{Milky Way Project bubbles}

The Milky Way Project (MWP\footnote{http://www.milkywayproject.org}), a web-based citizen science project created by the Zooniverse\footnote{http://www.zooniverse.org}, invites users to classify bubble-shaped structures in the ISM in images from the Spitzer GLIMPSE and MIPSGAL surveys~\citep{Benjamin2003a, Carey2009a}. The method is based on the work performed by~\citet{Churchwell2006, Churchwell2007}: annular ellipses are drawn on the images with adjustable inner and outer axes and position angles. The raw classification data are then processed using a clustering algorithm to produce a final catalog. The project and first data release (DR1) is described in detail by~\citet{simpson_dr1}, and again in ~\citet{Kendrew2012}; we refer to these publications for background, and focus on the relevant points for our analysis here only.

\begin{figure}
    \includegraphics[width=8cm]{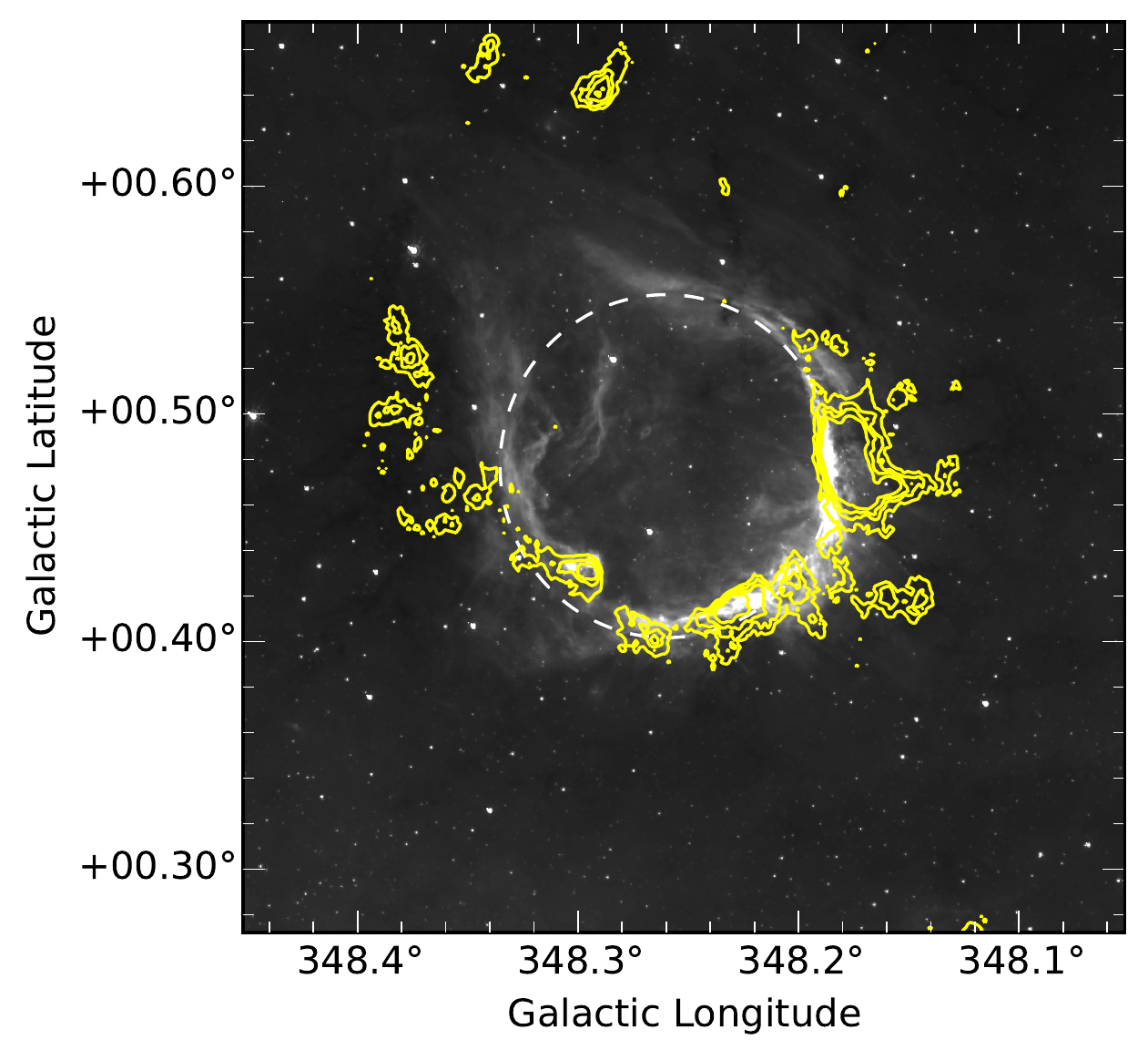}
    \caption{The ``perfect bubble'' RCW120~\citep{Deharveng2009} at 8~\um~(Spitzer IRAC band 4) in greyscale image. The contours show the 870~\um~continuum flux from the ATLASGAL survey map; contour levels represent approximately 3, 5, 7 and 10-$\sigma$ levels. The dashed circle shows the bubble location and size based on the MWP volunteer classifications, yielding an~\reff~of 4.52\arcmin. }\label{fig:rcw120}
\end{figure}

The MWP-DR1 catalog contains two separate types of bubbles: the small and large bubbles, denoted MWP-S and MWP-L, respectively. The MWP-S set represents bubbles that are not well resolved in the images presented to the users. They were classified using a simplified toolset, and as a result do not have full ellipticity information associated with them. In ~\citet{Kendrew2012} we showed how some of the limitations of the analysis were related to uncertainty on bubble locations and the bubbles' evolutionary stage; for this reason the present work uses only the statistically more robust MWP-L catalog (3744 bubbles). For illustration, Figure~\ref{fig:rcw120} shows the~\hiir~RCW120, referred to as ``the perfect bubble''~\citep{Deharveng2009} and denoted MWP1G348260+004774 and S7 in the MWP-DR1 and~\citet{Churchwell2006} catalogs respectively, as imaged by Spitzer at 8~\um. 
Within the common area of coverage with the ATLASGAL survey ($|l| \leq 60^{\circ}$, $|b| \leq 1^{\circ}$) we find \nbubs~bubbles. We note that the bubble catalog represents a heterogeneous sample of objects, containing~\hiirs and a small number of supernova remnants, evolved stellar bubbles and spurious detections.~\citet{Beaumont2014} improve on the detection of classical~\hiirs~using a machine learning algorithm; however for consistency with~\citet{Kendrew2012} we use here the full MWP-L catalog from MWP-DR1.s

\begin{figure}
    \includegraphics[width=8cm]{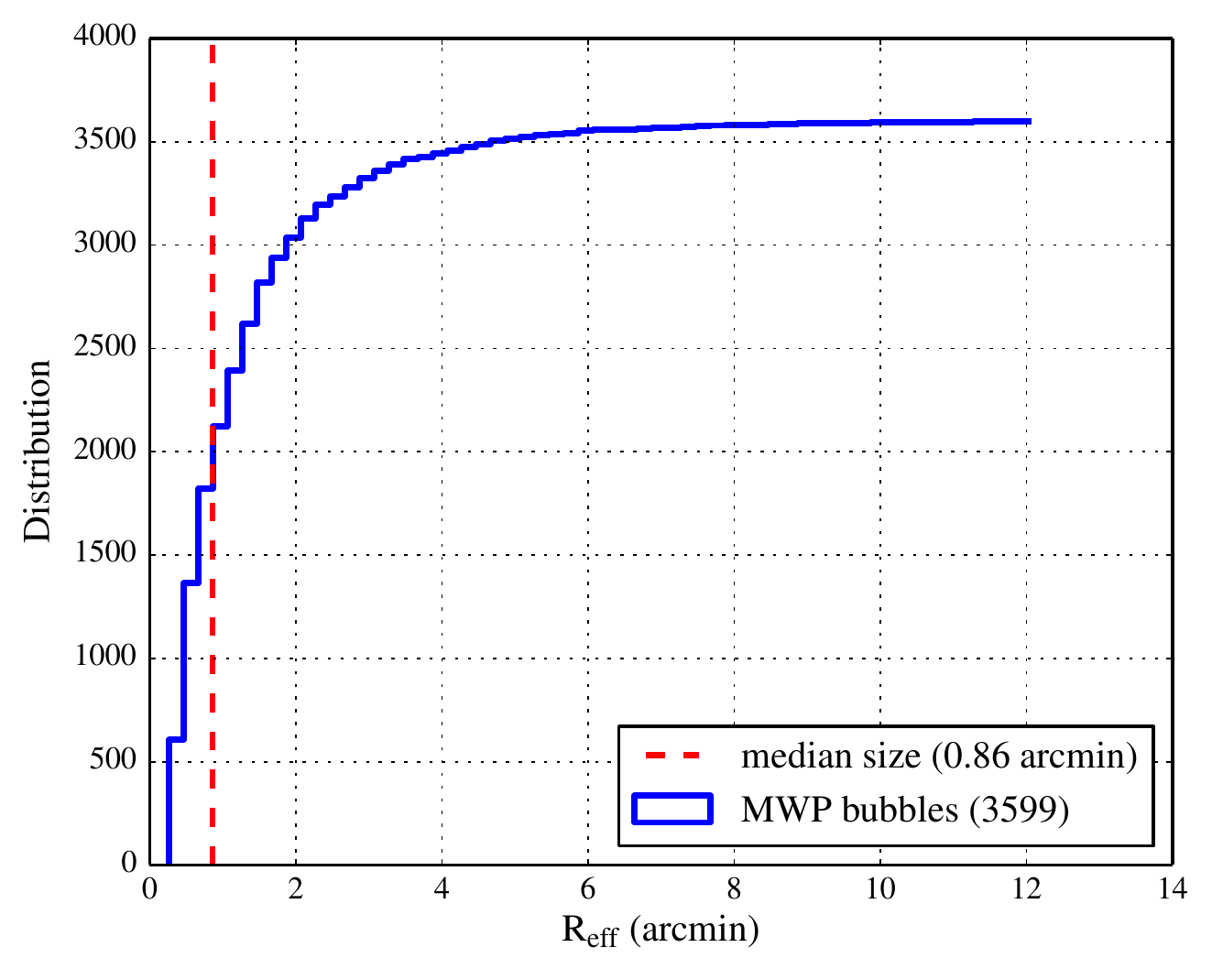}
    \caption{Cumulative distribution of effective radii of the sample of MWP bubbles used in the analysis.}\label{fig:mwp_rdist}
\end{figure}

As in ~\citet{Kendrew2012}, the key metric we employ in this paper is the bubbles' effective radius (R$_{\rm eff}$), defined as an average of the major and minor ellipse axes:

\begin{equation}
    R_{\rm eff}=\frac{(R_{out}r_{out})^{0.5}+(R_{in}r_{in})^{0.5}}{2}\label{eq:reff}
\end{equation}

where R$_{out}$, R$_{in}$ are the outer and inner semi-major axes of the bubble's best-fit annular ellipse shape, and r$_{out}$, r$_{in}$ the outer and inner semi-minor axes, respectively. These values are listed for each bubble in the DR1 catalog. The distribution of effective radii of the 3599 MWP bubbles in the sample used in this work is shown in Fig.~\ref{fig:mwp_rdist}.

In contrast with ~\citet{Kendrew2012}, the current analysis also includes the region $-10\degree \leq$ l $\leq 10\degree$, as the catalog used for correlation also covers this range (the RMS catalog does not). In Figures~\ref{fig:londist} and~\ref{fig:latdist} we show the distribution of the MWP catalog with Galactic longitude and latitude for reference, compared with the corresponding ATLASGAL source distribution. 

\subsection{The APEX Telescope Large Area Survey of the Galaxy (ATLASGAL)}

The ATLASGAL survey~\citep{Schuller2009} uses the LABOCA bolometer array on the Atacama Pathfinder Experiment (APEX) to map over 400 square degrees of the inner Galaxy at 870~\micron. The survey was carried out in multiple stages between 2007 and 2010, with the final survey area extending to $-60\degree \leq$ l $\leq +60\degree$ and $-1.5\degree \leq$ b $\leq +1.5\degree$. The APEX beam size is 19.\arcsec2. The full survey strategy and data reduction procedure is described in~\citet{Schuller2009}. We use here the ATLASGAL consortium's source catalog described in detail by \citet{Csengeri2014}.

The root mean square (rms) noise and thus the survey sensitivity varies across the survey area with galactic longitude; the rms noise ranges from 50 to 110 mJy beam$^{-1}$. \citet{Csengeri2014} adopt a survey-average value of 70 mJy beam$^{-1}$ and estimate a 5-$\sigma$ completeness limit of $\sim$ 10$^3$~\msun~to 20 kpc, or $\sim$40~\msun~to 4 kpc. We therefore assume that the majority of cold clumps with sufficient mass to form massive stars and clusters in the inner Galaxy are detected by the survey.

A spectroscopic follow-up study of the ATLASGAL data was performed by~\citet{Wienen2012}, who carried out observations of the NH$_3$ (1,1), (2,2) and (3,3) inversion transitions around 24 GHz with the Effelsberg 100-m telescope. They observed a flux-limited sample of 862 clumps in 5$\degree$ $\leq$ l $\leq$ 60$\degree$, $|b| \leq$ 1.5$\degree$, with a spectral resolution of 0.5 km s$^{-1}$. The spectra were used to measure velocities, linewidths, derive kinetic temperatures, and determine distances for the subsample. Of the 862 clumps, lines were detected for 750, of which 725 lie within the overlapping area with the MWP bubbles. Twenty of these clumps do not have reliable temperature measurements (T$_{kin} = 0$ K in the catalogue), and we exclude these sources from any temperature calculations.

We note that the clumps observed in the \citet{Wienen2012} study were extracted from the ATLASGAL maps using a similar but not identical source extraction procedure, from an older source catalogue than that provided in \citet{Csengeri2014}. Whilst the catalogues are highly consistent with each other in source extraction methodology, the \citet{Wienen2012} sources do therefore not necessarily have direct counterparts in the \citet{Csengeri2014} source catalogue. The \citet{Wienen2012} sources were selected to be compact ($<$ 50\arcsec) and $\geq$ 0.4 Jy beam$^{-1}$ in peak flux density, which the authors take to be representative of the ATLASGAL source sample. The \citet{Csengeri2014} catalogue contains 3363 sources in the region covered by the \citet{Wienen2012} survey; the \citet{Wienen2012} dataset is therefore representative but not complete. Furthermore, a peak flux of 0.4 Jy beam$^{-1}$ represents the 27th percentile in the full \citet{Csengeri2014} catalog, i.e. it contains some 2750 clumps fainter than this level, which is a significant number. We conclude that the \citet{Wienen2012} sample is biased towards brighter sources, and hence higher column densities.

At 870~\micron~the emission is dominated by thermal continuum radiation from cold dust. As the emission is optically thin, the H$_2$ column density as measured from the sub-millimeter source flux is directly proportional to the measured flux, and independent of distance.

\begin{equation}
    N(H_2) = \frac{F_{\nu}R}{B_\nu(T_D) \Omega \kappa_{\nu} \mu m_H}
    \label{eq:coldens}
\end{equation}

where $F_{\nu}$ is the flux density, $B_\nu(T_D)$ the blackbody flux at dust temperature $T_D$, $\Omega$ the beam solid angle, $\kappa_{\nu}$ the dust absorption coefficient, $R$ the dust-to-gas ratio, $\mu$ the mean molecular weight of the ISM (assumed to be 2.8), and $m_H$ the mass of a hydrogen atom. The H$_2$ column density was computed for all sources in the \citet{Csengeri2014} catalog, assuming a dust temperature $T_D$ of 20K and $\kappa_{\nu}$ of 1.85 cm$^2$ g$^{-1}$ (following~\citet{Csengeri2014}) and a constant $R$ of 100. We use for $F_{\nu}$ the peak flux from the \citet{Csengeri2014} catalog and obtain $\Omega$ from the beam FWHM of 19.2\arcsec, effectively calculating the beam-averaged column density for each source. The assumption of a single temperature of 20 K for all clumps can be tested by examining the distribution of kinetic temperatures obtained from the NH$_3$ spectroscopic observations by \citet{Wienen2012}. These show a range from approx. 15 to 50 K, with roughly 85\% of sources showing a $T_{kin}$ between 15 and 30K. Ignoring further underlying (significant) uncertainties, e.g. on the dust opacity, this translates to a factor of 2 in $N(H_2)$, we estimate this as the typical uncertainty on our calculated values. Given the high column densities required for massive star formation, all early evolutionary stages are associated with strong dust continuum emission. As such, the ATLASGAL data provide an excellent unbiased and systematic view on massive star formation over a range of evolutionary phases. 

\begin{figure*}[t]
    \centering
    \includegraphics{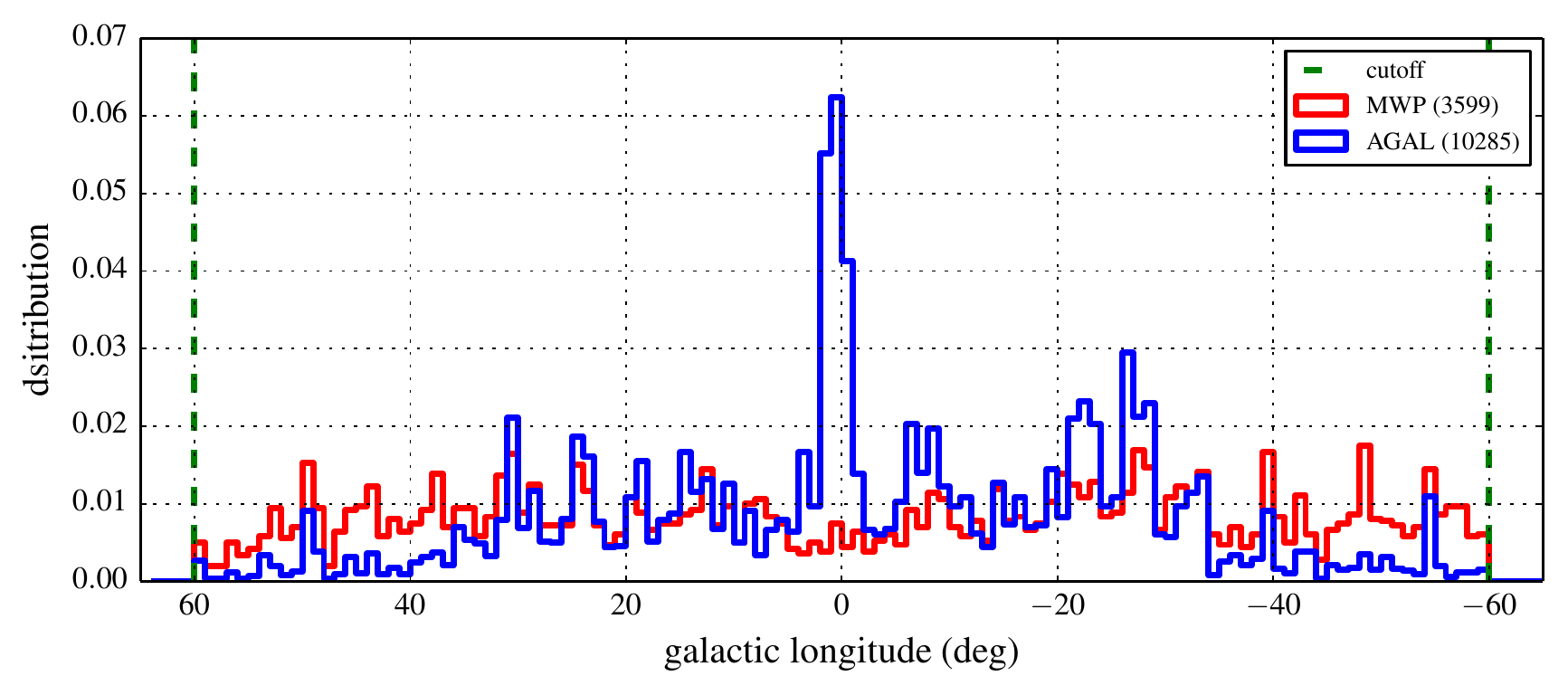}
    \caption{Distribution of MWP bubbles and ATLASGAL sources with galactic longitude. Counts are normalised; the absolute source number for each catalogue is indicated in the legend.}\label{fig:londist}
\end{figure*}

\begin{figure}[t]
    \includegraphics{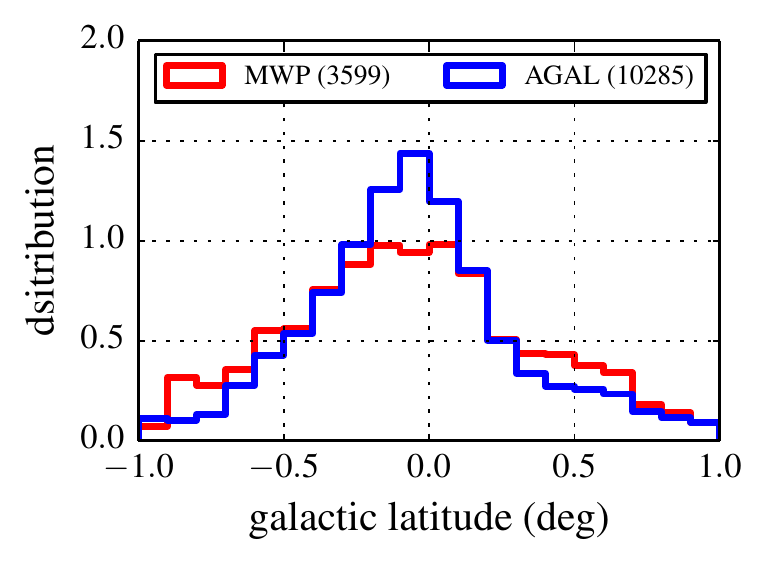}
    \caption{Distribution of MWP bubbles and ATLASGAL sources with galactic latitude. Counts are normalised; the absolute source number for catalogue is indicated in the legend.}\label{fig:latdist}
\end{figure}

\begin{figure}[h]
    \includegraphics{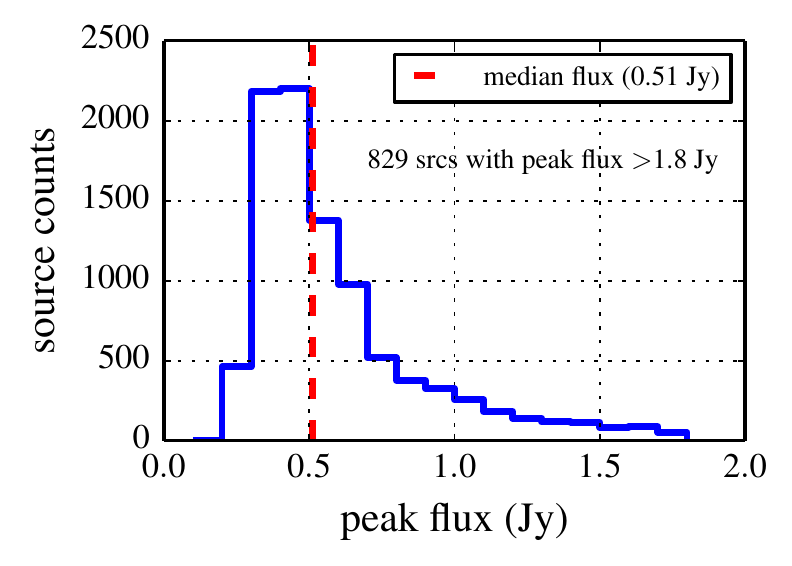}
    \caption{Distribution of peak fluxes for the~\nclumps~ATLASGAL sources in the catalog used. For clarity the plot shows only up to 1.8 Jy; approx. 8\% of sources have fluxes higher than this value.}\label{fig:agal_maxflux}
\end{figure}

The ATLASGAL source catalogue used for this work contains 10952 clumps and is described in detail by \citet{Csengeri2014}; within the common area with the MWP bubbles we find~\nclumps~clumps. Sources were extracted using the \emph{Multi-resolution and Gaussclumps} (MRE-GCL) method~\citep{Motte2007}, and are estimated to be complete to 97\% above 5-$\sigma$ and $>$99\% above 7-$\sigma$. The common coverage area with the MWP catalogue contains 10582 sources, for which the catalogue lists source size (FWHM in x and y, beam-averaged FWHM), peak and integrated fluxes. In our Figure~\ref{fig:rcw120} showing the bubble associated with~\hiir~RCW120 we show the ATLASGAL contour map in the vicinity of the bubble as illustration. The distribution of ATLASGAL sources in galactic longitude and latitude is shown in Figures~\ref{fig:londist} and~\ref{fig:latdist}, compared with the MWP bubbles. We show the distributions in peak fluxes in Figures~\ref{fig:agal_maxflux}. 

The longitude distribution of both samples shows a good correlation, broadly tracing the structure of the Milky Way Galaxy~\citep{simpson_dr1, Beuther2012}. Towards the central regions of the Galaxy the strong peak seen in the submillimeter clumps is absent from the observed bubble distribution. We identify three potential reasons for this discrepancy. First, the strong 24~\micron~background seen towards the Galactic Center likely masks the foreground bubbles, making them harder to detect visually by the MWP volunteers. Second, the formation of bubbles may be less efficient in this region, and their lifetimes shorter than in the disk, due to the increased pressure and turbulence in the Galactic Center molecular clouds. A possible third reason is the reduced star formation rate in the Central Molecular Zone proposed by~\citet{Longmore2013}; in this regard the lower bubble count is consistent with the observed lower counts in young stellar objects~\citep{Beuther2012}.

\section{Correlation analyses for clustering studies}\label{sec:corr_theory}

Angular correlation functions are commonly used in astrophysics to investigate statistically the clustering properties of a dataset, or of one dataset with regards to another. In ~\citet{Kendrew2012} we showed the use of such functions for the study of massive YSOs near infrared bubbles. The method used to study the statistical clustering of cold clumps near infrared bubbles uses the Landy-Szalay correlation function estimator~\citep{landyszalay93}, generalised for two different datasets~\citep{Bradshaw2011}:

\begin{equation}
	w(\theta)=\frac{N_{D_1D_2}(\theta)-N_{D_1R_2}(\theta)-N_{R_1D_2}(\theta)+N_{R_1R_2}(\theta)}{N_{R_1R_2}(\theta)}
\label{eq:xcorr}
\end{equation}

where $w(\theta)$ is the correlation as a function of angular distance $\theta$, $N$ represents the number of pairs between ``true'' data (subscript $D$) or random catalogs (subscript $R$). In our analysis we express the angular separation $\theta$ as a function of bubble $R_{\rm eff}$ to study specifically where overdensities are present with respect to the bubble rim. The pair counts were normalised to account for different catalog sizes. In essence, when applied to the MWP bubbles and ATLASGAL cold clumps, this calculation shows the density of cold dust clumps near infrared bubbles \emph{in excess} of what is expected from random distributions of bubbles and clumps.

To account for sampling errors we introduce a second level of randomisation: as well as generating entirely randomly distributed catalogs of bubbles and clumps, we implement a bootstrap resampling method. For each of N bootstrap iterations, a different random sampling is taken from the input data. Each randomly sampled instance of the catalogue maintains the number of sources as the input catalogue but allows for replacements.

The method is described in detail in ~\citet{Kendrew2012}, along with a number of methodological tests, which we do not repeat here. However a number of changes were implemented since ~\citet{Kendrew2012} to generalise the code to the current datasets. First, the generation of random (subscript R in equation~\ref{eq:xcorr}) catalogs was adapted to account for the strong peak in ATLASGAL sources towards $l = 0\degree$. Instead of generating a uniform distribution in galactic longitude, source counts were matched in bins of $\Delta l = 5\degree$. As before, the random latitudes were generated from the best-fit Gaussian to the input data latitudes, and the random bubble R$_{\rm eff}$ values are drawn from the best-fit lognormal distribution of the data values.

Second, the methods used to perform computationally intensive work in the code were adapted to cope with the much larger sample size of the ATLASGAL data catalog. With random catalogs typically 50 times larger than the data catalogs, catalog generation and pair count operations become prohibitively slow unless fast computational methods are implemented. The generation of random catalogs was vectorised to avoid excessive loop operations. The pair counts calculation now includes a K-dimensional (KD)-tree algorithm to compute efficient distance matrices. As before, the code is publicly available\footnote{http://www.github.com/skendrew}.

Given a large enough sample size, the correlation functions can be ranked along additional criteria. In ~\citet{Kendrew2012} we ranked the bubbles in size bins and studied the correlation of MYSOs near bubbles as a function of size. Given the larger size of the \citet{Csengeri2014} ATLASGAL catalog, such a ranking can now also be performed in the clump dataset.

These code updates were extensively tested for compatibility with previous results. They do not affect the validity of the preliminary method tests performed in ~\citet{Kendrew2012}.

\section{Results}

\subsection{Associated and control samples}

As a first test, simple distance calculations were performed to identify the ATLASGAL clumps that are associated with a bubble. The ``association'' is defined as an angular separation of $<$ 2 bubble effective radii (R$_{\rm eff}$; see Equation~\ref{eq:reff}) from the nearest bubble, as in ~\citet{Kendrew2012}. As this computation is performed in 2-D space, it does not take line-of-sight confusion into account, i.e. some of the clumps that appear near bubbles may be located in the foreground or background of the bubble. For the full \citet{Csengeri2014} sample, we find 4961 clumps to be associated with a bubble, or 48\% of our sample. 26\% of the total (2736 clumps) lie specifically near a bubble rim (defined as 0.8 R$_{\rm eff}< \theta <$ 1.6 R$_{\rm eff}$). We define a ``control'' sample to consist of those clumps that lie $>$ 3 R$_{\rm eff}$ from the nearest bubble. This sample contains 35\% of all cold clumps (3623 clumps). Uncertainties on these numbers were calculated with a bootstrap resampling method with replacements, where we perform the clump-bubble proximity calculation with a different randomly generated selection of clumps. The observed standard deviation on the bubble-associated and control clump counts over 100 iterations is 1-2\%. 

\begin{figure}
    \includegraphics[width=9cm]{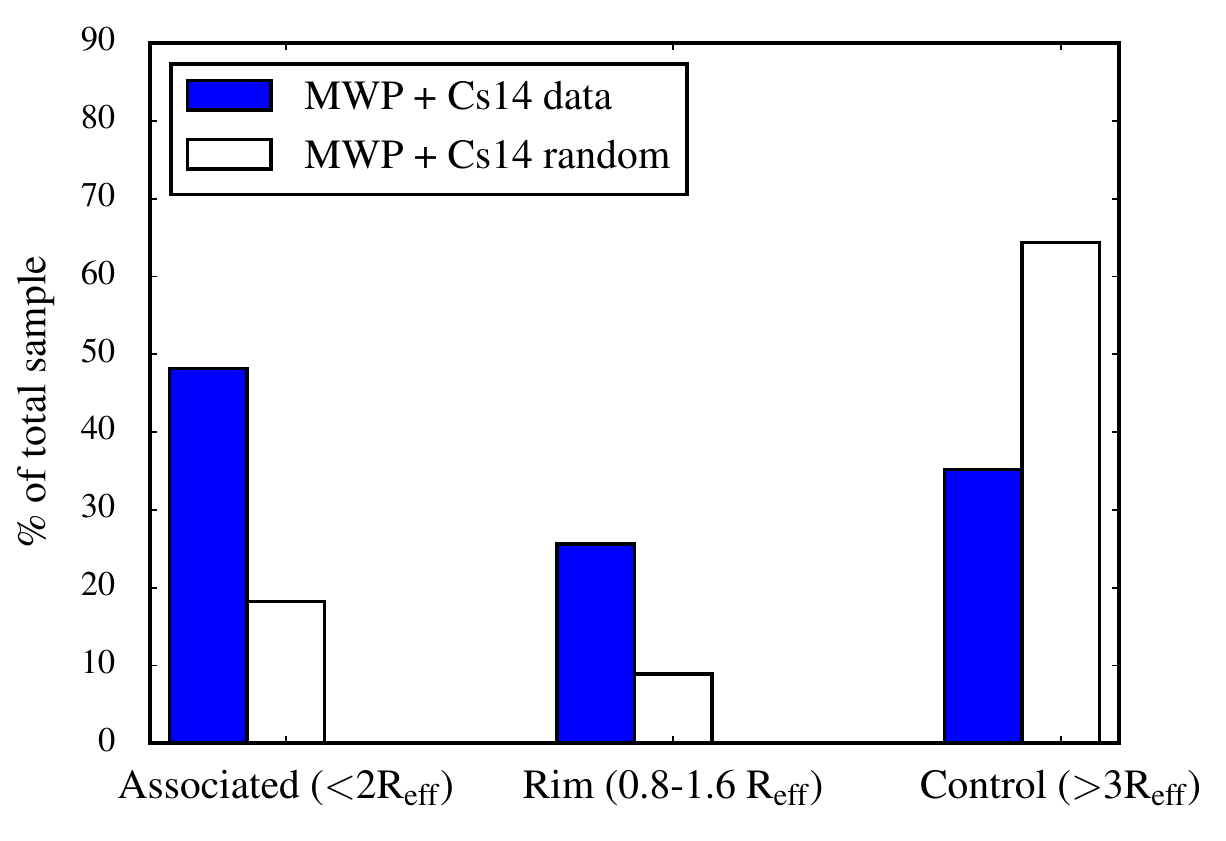}
    \caption{Bar plot showing the percentages of ATLASGAL cold clumps (from~\citet{Csengeri2014} (Cs14)) associated with (rims of) MWP bubbles, and those in the control sample. The filled bars show the associations for the data, open bars those for the randomly generated catalogues. Typical error bars are 1-2\%.}\label{fig:divsamp_bars}
\end{figure}

To test the significance of these numbers compared with what is expected from random distributions of bubbles and clumps on the sky, the same calculation was performed with randomly generated catalogs of both object types of the same size as the ``true'' data catalogs. The generation of these random catalogs follows closely the procedure described in Section~\ref{sec:corr_theory}: number counts match in $\Delta l$ bins of 5$\degree$, latitudes are drawn from the best-fit Gaussian distribution, and bubble R$_{\rm eff}$ from the best-fit lognormal distribution of the input data. Uncertainty on the number counts of associated and control samples was obtained by performing this calculation on 1000 separate randomizations. For these random distributions of bubbles and clumps, we find that 18 $\pm$ 1\% of clumps are located within 2 R$_{\rm eff}$ from the nearest bubble, with 9 $\pm$ 1\% found near a bubble rim. 64 $\pm$ 2\% of clumps are found in the field. The results of these data and random calculations are visualised in Figure~\ref{fig:divsamp_bars}.

Performing the same tests on the smaller \citet{Wienen2012} sample, for which NH$_3$ spectroscopic data are available, yield similar numbers for both the data and the randomised test catalog: 55\%, 31\% and 30\% for associated, rim, and control samples, respectively. Association counts in randomised catalogs of the same size as the \citet{Wienen2012} sample (725) were consistent with those over the full coordinate range. The numbers suggest that while the number of expected chance alignments is non-negligible, the level of association between MWP bubbles and ATLASGAL clumps is significant. The number of clumps found near a bubble rim, in particular, is a factor 3 higher than what is expected from random distributions.

\subsection{Physical properties of bubble-associated vs. field clumps}

The bubble-associated and control sub-samples of the catalog can be examined more closely for differences in physical properties, which gives an insight into the physical environment near bubbles and the possible effect the energy of the expanding bubble is having on its surrounding medium. From the \citet{Csengeri2014} ATLASGAL catalog we can compare sizes and and peak fluxes of the associated and control clumps. Median beam-convolved FWHM are 27.7\arcsec~and 27.6\arcsec for the bubble-associated and control clumps, respectively. This quantity is defined as follows in \citet{Csengeri2014}:

\begin{equation}
    FWHM = \sqrt{\Theta_{\rm maj} \times \Theta_{\rm min}}
\end{equation}

where $\Theta_{\rm maj}$ and $\Theta_{\rm min}$ are are the beam-convolved major and minor axes of the 2-D Gaussian fits to the clumps.

\begin{figure}
    \includegraphics{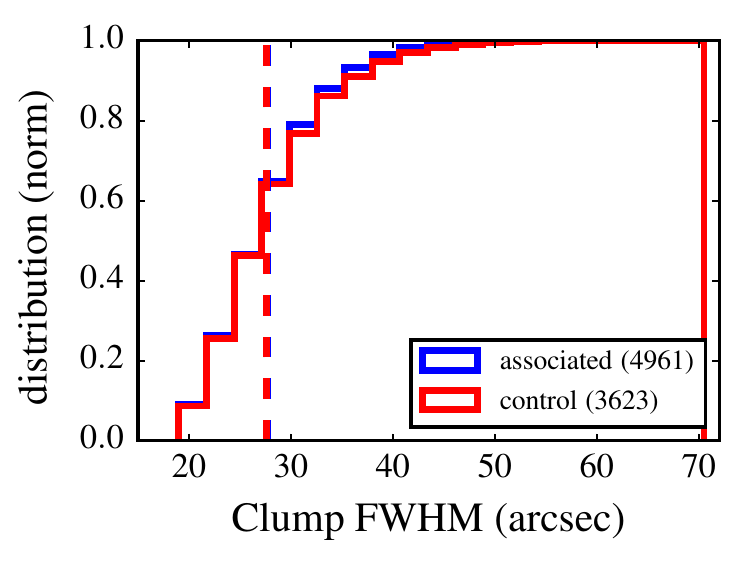}
    \caption{Distribution of ATLASGAL clump beam-convolved FWHM, in arcsec, comparing the bubble-associated sample (blue) with the control clumps (red). Vertical dashed lines indicate the median values for each sample: 27.6\arcsec~and 27.7\arcsec~for control and associated clumps, respectively (see text). The distributions are statistically identical.}\label{fig:agal_divsamp_reff}
\end{figure}

The size distributions for these samples are shown in Figure~\ref{fig:agal_divsamp_reff}, which shows them to be virtually identical. A k-sample Anderson-Darling (A-D) test~\citep{Scholz1987} does not provide strong evidence for the distributions being statistically different\footnote{We note that in~\citet{Kendrew2012} we use the 2-sample Kolmogorov-Smirnov (K-S) test to check the similarity of distributions. The A-D test more sensitive to differences in the tails of the distributions, e.g.~\citet{Hou2009}; we therefore use the A-D test throughout this work. We did however compare to the outcome of the 2-sample K-S test on the same distributions and found the results to be consistent from both tests. The change in statistical tests with our previous work does not change any findings}. Cold dense clumps near bubbles thus do not differ in size from their counterparts in the field.

\begin{figure}
    \includegraphics{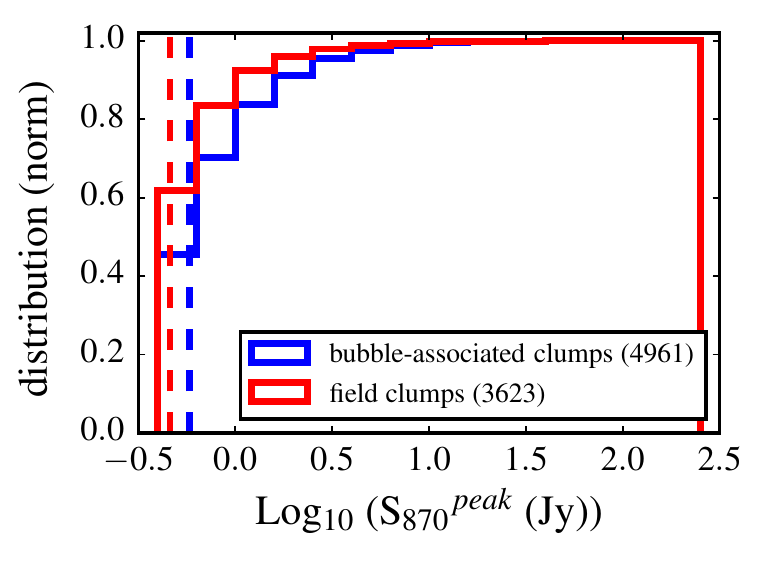}
    \caption{Cumulative distribution of ATLASGAL 870~\micron~peak fluxes, in Jy, comparing the bubble-associated sample (blue) with control clumps (red). Vertical dashed lines give the median values, 0.58 Jy and 0.46 Jy for associated and control samples, respectively.}\label{fig:agal_divsamp_pflux}
\end{figure}

Comparison of the peak 870~\micron~fluxes of the ATLASGAL clump samples is shown in Figure~\ref{fig:agal_divsamp_pflux}. The median value of the peak fluxes of the bubble-associated clumps is found to be higher than that of the field clumps: 0.58 Jy beam$^{-1}$ versus 0.46 Jy beam$^{-1}$, respectively. The k-sample A-D test rules out sampling effects as cause of the different distributions to a high level of confidence. Following equation~\ref{eq:coldens}, the flux is related to the clump column density independently of distance, for a given dust temperature, beam size, dust-to-gas ratio and dust absorption coefficient. Assuming these quantities do not vary systematically with distance in the Galactic Plane, the different flux distributions shown in Figure~\ref{fig:agal_divsamp_pflux} imply that cold dust clumps near infrared bubbles have statistically higher peak column densities than those in the field. 

\begin{figure*}
    \includegraphics{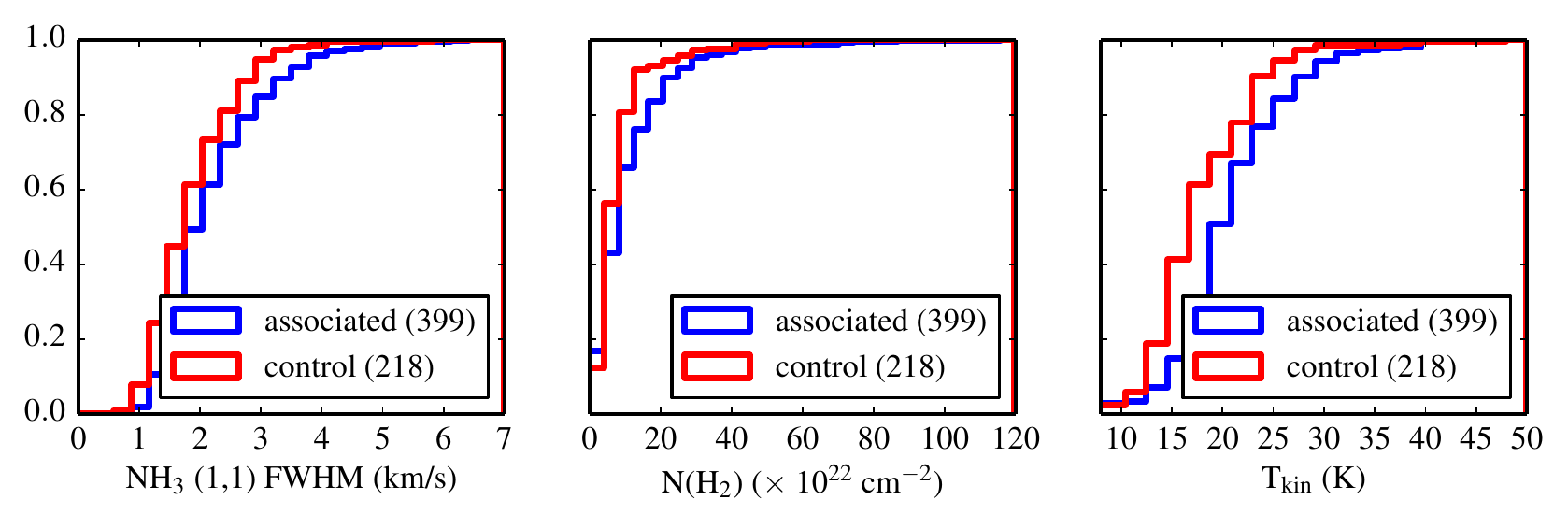}
    \caption{Cumulative distribution of ATLASGAL clumps from the \citet{Wienen2012} (W12) sample with NH$_3$ spectroscopic observations, comparing bubble-associated and control clumps. L: NH$_3$ (1,1) FWHM linewidth; M: H$_2$ column density; R: kinetic temperature. The plots show how dust clumps near bubbles have systematically larger linewidths, H$_2$ column densities and kinetic temperatures.}\label{fig:nh3_divsamp}
\end{figure*}

The \citet{Wienen2012} sample of clumps lends itself to further comparisons of physical properties between bubble-associated and control clumps. In Figure~\ref{fig:nh3_divsamp} we compare these sub-samples in NH$_3$ (1,1) linewidth, H$_2$ column density and kinetic temperature. In all three cases, the clumps near bubbles show higher values and distributions that are statistically distinct from their counterparts in the field, the A-D test ruling out the similarity of the distributions to $<< 1\%$.  Median linewidths of the (1,1) line are found to be 2.06 and 1.83 km s$^{-1}$ for bubble-associated and control clumps, respectively. Given typical temperatures of 20 K, the linewidths of all clumps in the sample are dominated by non-thermal contributions ($\Delta v_{\rm thermal}$ at 20 K $\sim$ 0.2 km s$^{-1}$). 

Given the small differences in median values we perform a further test to investigate the significance of the differences between the distributions, for peak flux of the \citet{Csengeri2014} clumps and NH$_3$ (1,1) linewidths for the \citet{Wienen2012} sample. These are directly measured quantities for which uncertainties are well understood. For each of these quantities we calculate the 25th, 50th, 75th and 90th percentile values for the bubble-associated and control clumps. We then calculate the significance (the signal to noise, `S/N') of the difference between the two distributions as a function of the 1-$\sigma$ noise for the given quantity. 

\begin{figure}
    \includegraphics{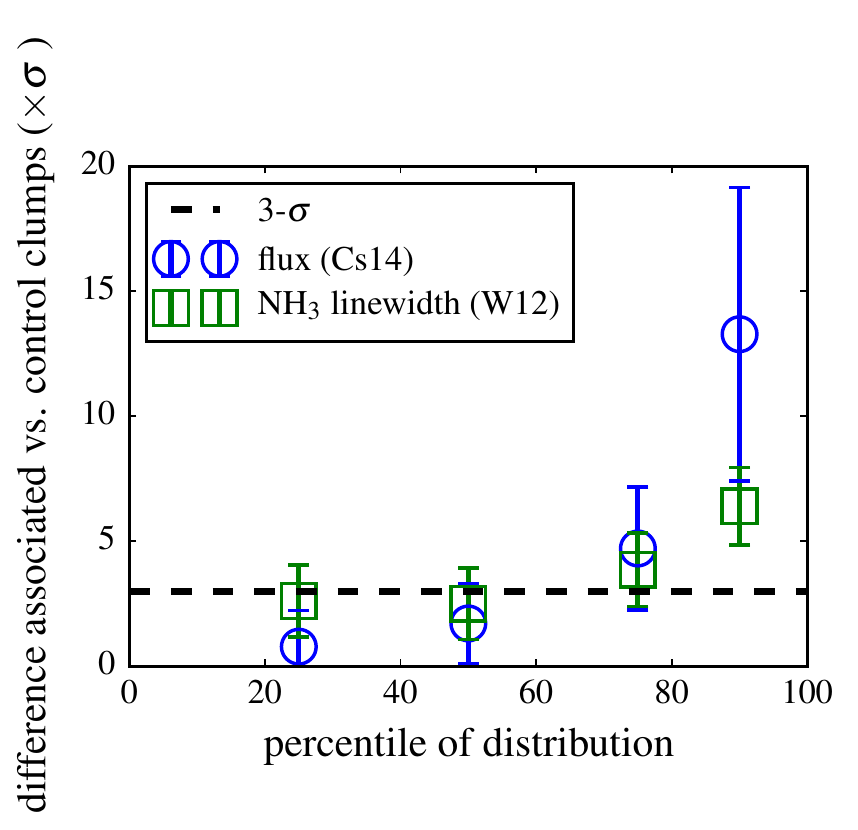}
    \caption{Significance of the difference between distributions of bubble-associated and control clump samples, in peak fluxes in the \citet{Csengeri2014} clump sample (blue circles) and in NH$_3$ (1,1) linewidth from the \citet{Wienen2012} (W12) spectroscopic subsample. The increase in significance towards higher percentiles indicates that the distribution of the relevant quantity for the bubble-associated clumps is skewed towards higher values than that of the control clumps.}\label{fig:agal_divsamp_distdiff}
\end{figure}

\citet{Csengeri2014} give an rms noise level on the flux measurements of 0.07 Jy beam$^{-1}$ (varying with longitude between 0.05 and 0.12 Jy beam$^{-1}$). For the linewidths, we adopt a mean uncertainty of 0.09 km s$^{-1}$ on the measurements, based on the uncertainties reported in the catalogue. The uncertainty on the linewidths is similar for bubble-associated and control clumps, so adopting a single uncertainty value for the full sample is justified.  The S/N values for both quantities at the selected percentile levels are shown in Figure~\ref{fig:agal_divsamp_distdiff}. The figure shows that whilst the difference between the values for the 25th and 50th percentiles is at best marginally significant (at $\sim$ 2.5-3-$\sigma$), the significance increases towards the high end of the distribution. This indicates the distribution of linewidths and fluxes of the bubble-associated clumps has a larger high-end tail than that of the control clumps, consistent with the outcome of the A-D test that indicates the statistical difference of the distributions.

\subsection{Dust clump clustering near infrared bubbles}

\begin{figure}
    \includegraphics[width=10cm]{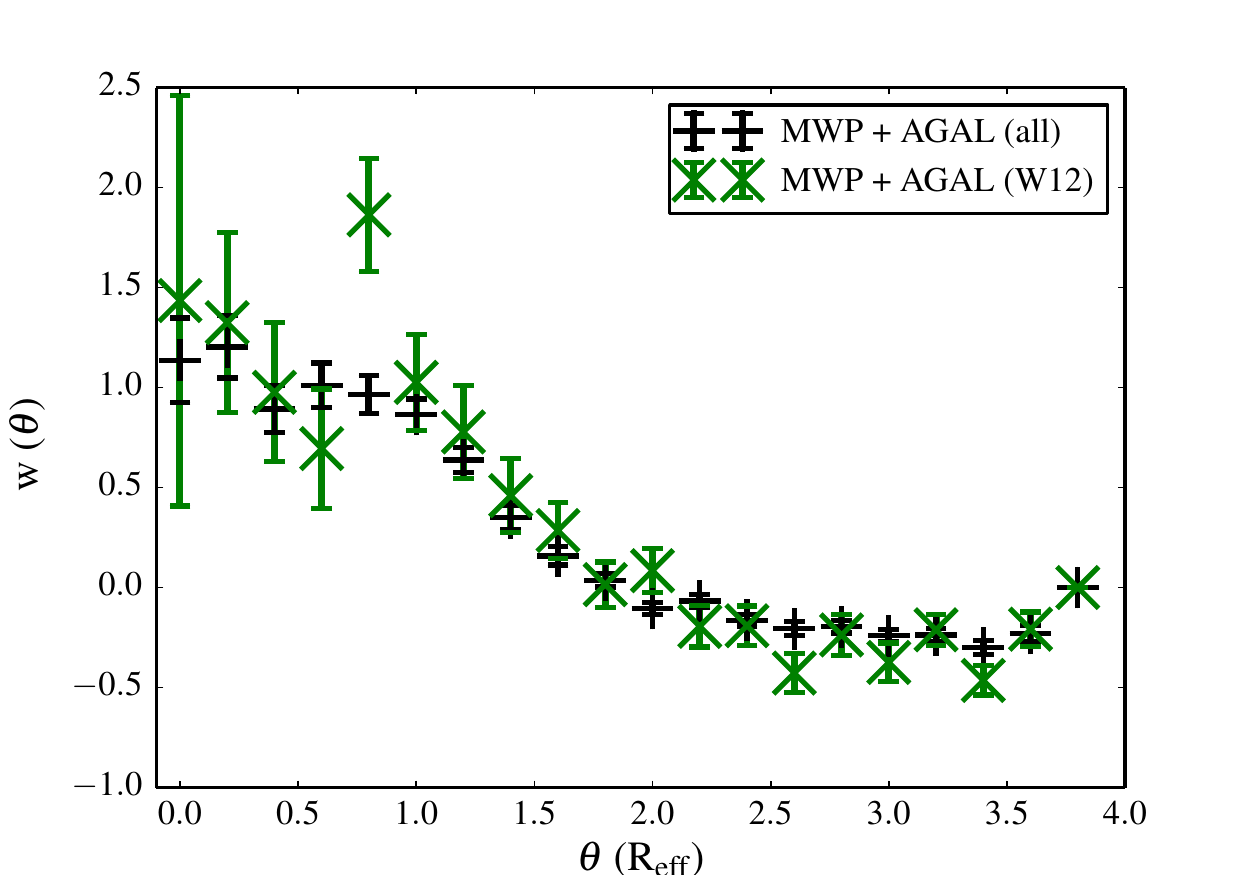}
    \caption{Angular correlation function of the ATLASGAL clumps near MWP bubbles, for the full sample from both catalogues, \citet{Csengeri2014} (`all') and \citet{Wienen2012} (`W12'). The randomly generated catalogues contained 20 times the number of data sources, and 100 bootstrap iterations were performed with replacement for uncertainty estimates. The correlation functions are broadly consistent with each other, the \citet{Wienen2012} points however showing larger error bars and increased scatter due to the smaller sample size.}\label{fig:corr_all}
    \end{figure}

Using the formalism presented in Section~\ref{sec:corr_theory}, we investigate the distribution of cold dust clumps near infrared bubbles statistically by calculating the angular correlation function between the datasets. In our application of the method, the MWP bubbles and ATLASGAL clumps represent datasets 1 and 2, respectively. The random catalogues were chosen to be 50 times larger than the input data, and 100 bootstrap iterations were performed for accurate uncertainty estimates. The angular correlation function $w(\theta)$ for the full bubble and clump catalogues is shown in Figure~\ref{fig:corr_all}, as well as the correlation between bubbles and the \citet{Wienen2012} clump sample. The function shows a marked increase towards bubble-clump separations of $<$ 1 \reff, indicating an overdensity of cold dust clumps towards and near infrared bubbles.

The correlation between MWP and the \citet{Wienen2012} limited longitude sample follows that of the full \citet{Csengeri2014} catalogue, with one marked outlier point in the 0.8-1~\reff bin. This suggests that in the 5$\degree < $ l $< 60\degree$ range we find a strong overdensity of dust clumps near the rims of bubbles. We note however that because of the smaller number counts in the \citet{Wienen2012} sample the uncertainty on the correlation is larger than in the sample of \citet{Csengeri2014} and more sensitive to outliers.

\subsubsection{Bubble sizes}\label{sec:bub_size}

\begin{figure}
    \includegraphics[width=9cm]{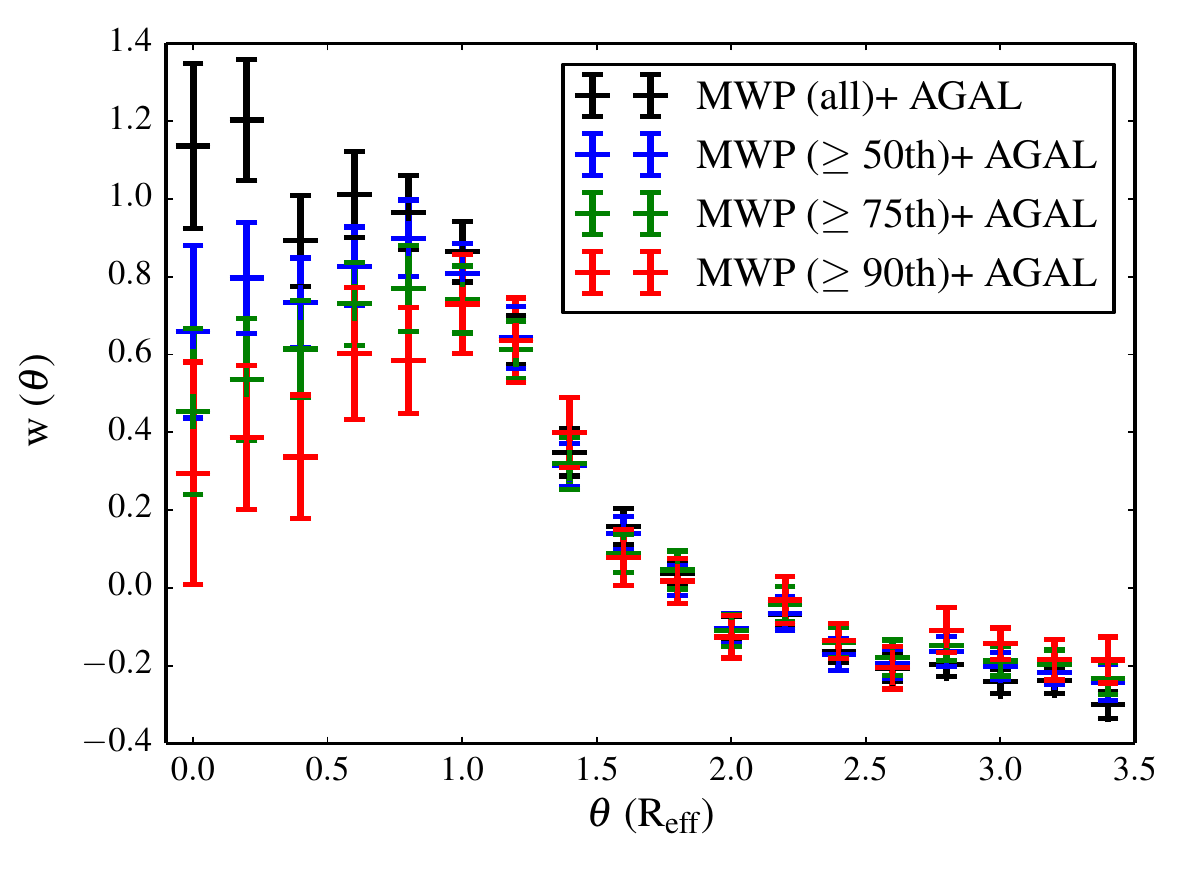}
    \caption{Correlation functions of MWP bubbles and ATLASGAL cold clumps, ranked by bubble sizes. Correlation near all bubbles is shown in black, and near bubbles larger than the 50th, 75th and 90th percentiles in blue, green and red, respectively. Larger bubbles show a reduced correlation with cold clumps towars their interiors compared with their smaller counterparts.}\label{fig:corr_sizes}
\end{figure}

\citet{Kendrew2012} showed an evolution in the correlation function with increasing bubble size, suggesting that the overdensity of star formation tracers along bubble rims increases with bubble size. We perform the same analysis with the ATLASGAL clumps, binning the bubbles into samples with \reff $\geq$ 0.87\arcmin, $\geq$ 1.57\arcmin and $\geq$ 2.72\arcmin, corresponding to the 50th, 75th and 90th percentiles of the full sample respectively. The results of this analysis are shown in Figure~\ref{fig:corr_sizes}. The observed correlations are almost identical for $\theta \geq$ 1~\reff. At smaller separations, i.e. projected towards the bubble interior, increasing bubble size is associated with a markedly lower correlation with cold clumps. We interpret this as the statistical signature of the bubbles' driving sources gradually clearing dense material from their interiors and sweeping material up along its rim. All bubbles show strong ($>$ 5-$\sigma$) overdensities of cold clumps around 1 bubble~\reff, where the swept-up material accumulates around the bubbles' expanding edge.

\subsubsection{Clump column density}

\begin{figure}
    \includegraphics[width=9cm]{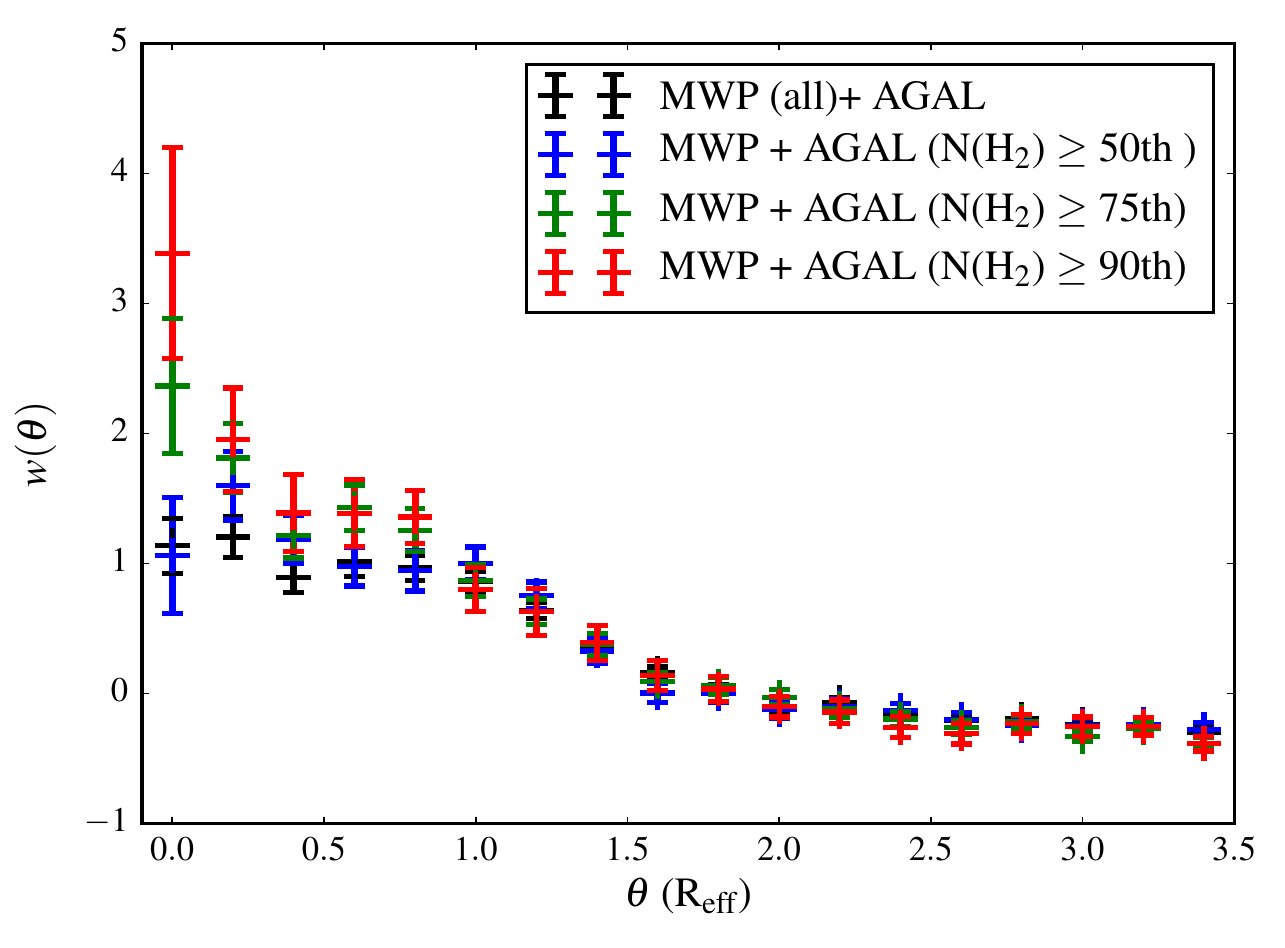}
    \caption{Correlation functions of MWP bubbles and ATLASGAL cold clumps, ranked by beam-averaged clump column densities. As in Figure~\ref{fig:corr_sizes}, the correlation function for the full samples is shown in black for reference. The blue, green and red points show the correlation between bubbles and clumps of column densities, N(H$_2$) $\geq 7 \times 10^{21}$ cm$^{-2}$ (50$^{th}$ percentile), N(H$_2$) $\geq 1 \times 10^{22}$ cm$^{-2}$ (75$^{th}$ percentile), N(H$_2$) $\geq 2 \times 10^{22}$ cm$^{-2}$ (90$^{th}$ percentile), respectively.}\label{fig:corr_nh2}
\end{figure}

The large number of clumps in the main~\citet{Csengeri2014} ATLASGAL catalog allows us to introduce additional rankings in the analysis that was not feasible with the RMS data in ~\citet{Kendrew2012}. The correlation analysis was repeated with the ATLASGAL clumps ranked in H$_2$ column density, as calculated from the catalog fluxes using equation~\ref{eq:coldens}. As with the bubble ranking shown in Section~\ref{sec:bub_size}, we extract clump samples with increasing sizes: clumps with N(H$_2$) $\geq 7 \times 10^{21}$ cm$^{-2}$ (50$^{th}$ percentile), N(H$_2$) $\geq 1 \times 10^{22}$ cm$^{-2}$ (75$^{th}$ percentile), N(H$_2$) $\geq 2 \times 10^{22}$ cm$^{-2}$ (90$^{th}$ percentile), and re-calculate the correlation function. The results are shown in Figure~\ref{fig:corr_nh2}, with the correlation between the full bubble and clump samples shown for comparison. Interestingly, we see a pronounced overdensity of high column density clumps towards bubble interiors, albeit with a higher scatter. The overdensity near the rim is also marginally raised over that seen in the full sample. We cannot say whether these clumps are located within the bubble, or outside the front/back surface, however given that the overdensity of these clumps near the rim is only slightly raised over that seen in the full sample, we propose that a sizeable fraction of them are located inside the bubble rim.

\subsection{ATLASGAL clump auto-correlation}

To study the intrinsic clustering of the cold dust clumps, we performed an auto-correlation analysis of the ATLASGAL cold clumps. This is required to rule out that any findings from the two-sample correlation analysis is due to intrinsic properties of the data. For the auto-correlation a random catalogue of clumps is generated using the same as described for the two-sample correlation function, with 20 times more sources than the input data. 100 bootstrap iterations are performed to estimate the uncertainty on the auto-correlation values, which were computed to 6\arcmin~from the clumps' centre. The clump-clump separation was in this case not expressed in units of clump radius, however we begin the computation at a minimum separation of 0.2\arcmin, which is approximately the beam half-width of the APEX telescope. 

\begin{figure}
    \includegraphics[width=9cm]{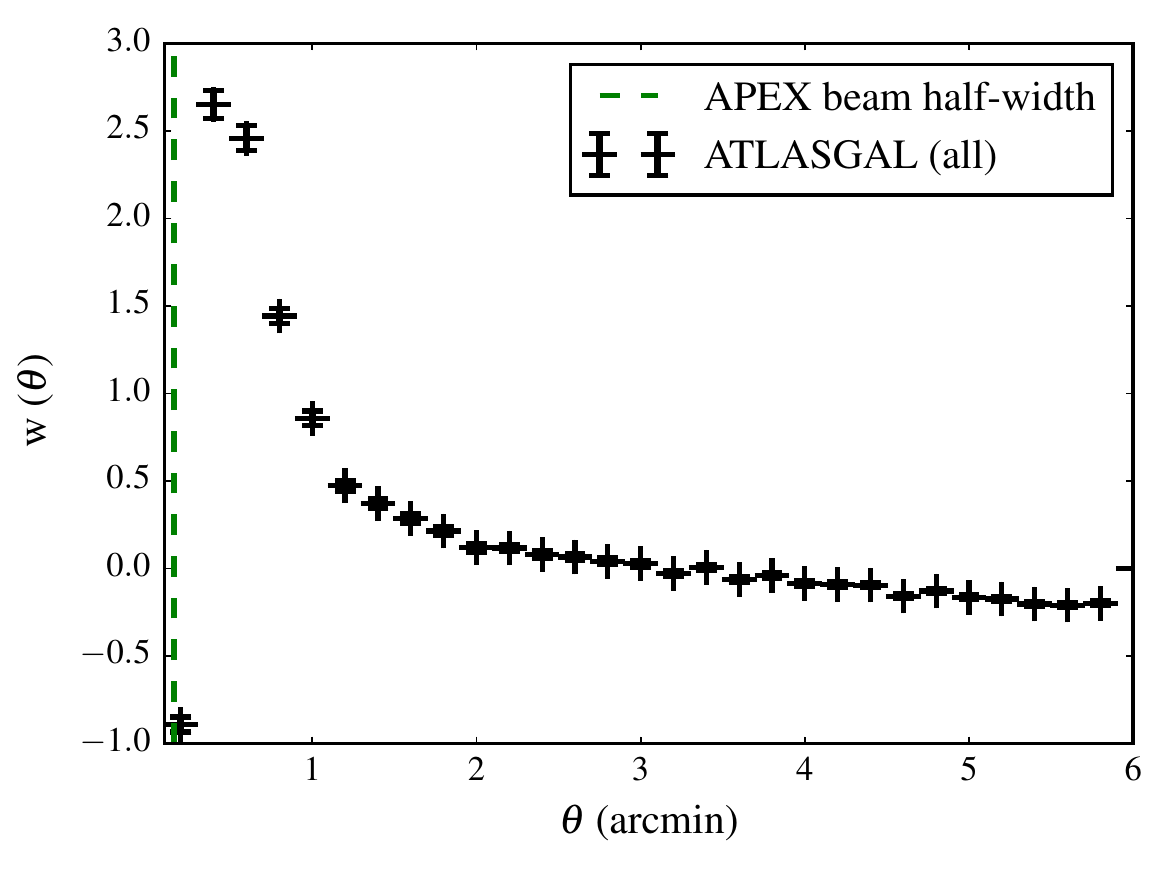}
    \caption{Angular auto-correlation of the ATLASGAL clumps. The vertical dashed line indicates the beam half-width of the APEX telescope. The clumps appear strongly clustered at small separations, as is expected.}\label{fig:c14_acorr}
\end{figure}

The results, shown in Figure~\ref{fig:c14_acorr}, indicate that the clumps are strongly clustered at short separations, peaking at the distance equivalent to the half-beam width. Given the clustered nature of star formation, this is an expected finding. The clustering decreases rapidly at separations larger than the beam, reaching a baseline level around 2\arcmin~where no excess clumps are found over the random distribution. Importantly, we do not see any systematic clustering on spatial scales equivalent to typical bubble sizes (e.g. the median bubble~\reff~of the 90$^{\rm th}$ percentile bubbles, 2.7\arcmin). This rules out that any overdensity seen in the two-sample correlation function reflects the intrinsic clustering of the dust clumps rather than a physically significant association.

\subsection{Clump physical conditions as a function of clump-bubble separation}

In the previous sections we have described the physical properties of dust clumps near infrared bubbles on the one hand, and quantified the overdensity of clumps towards and surrounding bubbles on the other. In this section we combine these measurements to compute statistically how the physical properties of the ATLASGAL clumps change as a function of distance from a bubble; in other words, we characterise the spatial scale over which the physical environment near IR bubbles differs from ``typical'' conditions in the field. These calculations are carried out using the \citet{Wienen2012} sample of clumps with NH$_3$ spectra, focusing in particular on the kinetic temperature of the gas and the observed linewidth of the (1,1) line. In this analysis we exclude the 20 clumps with T$_{\rm kin} = 0.0$ K, leaving 705 clumps and 1688 bubbles in the common survey area.

\begin{figure}
   \includegraphics[width=9cm]{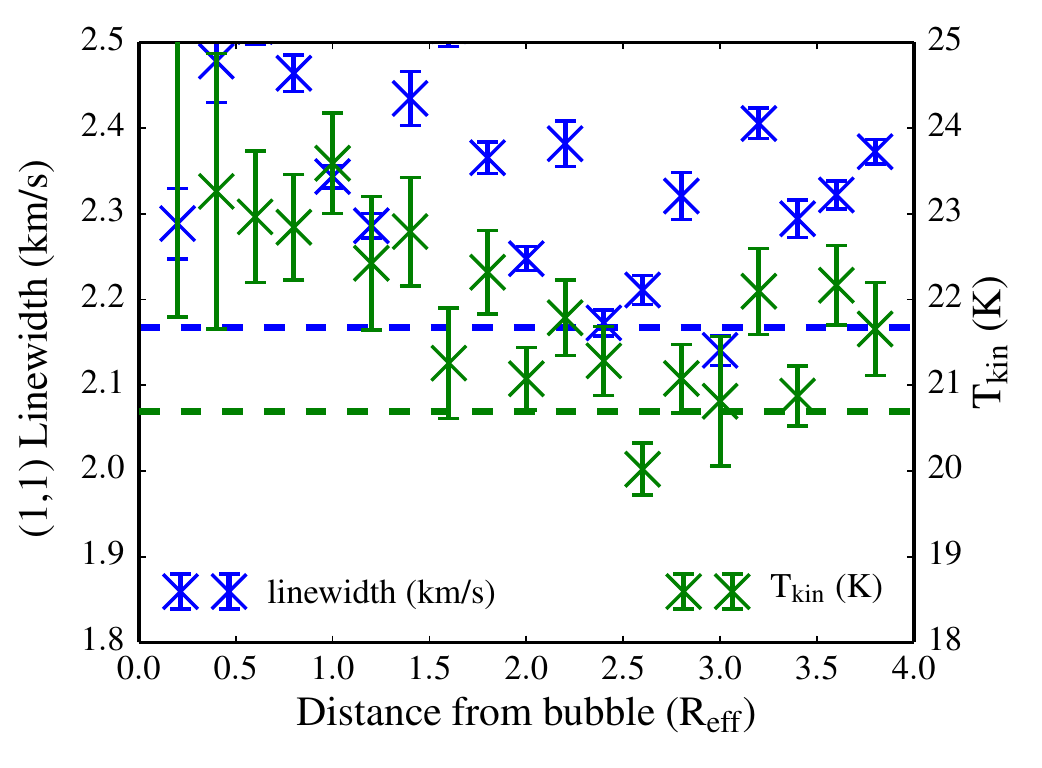}
    \caption{Kinetic temperature (right axis) and NH$_3$ (1,1) linewidth (left axis) of ATLASGAL clumps, as a function of radial distance from bubbles (in~\reff units; binsize = 0.2~\reff). The mean values of the overall sample is shown by the dotted lines. Errorbars represent 1-$\sigma$ uncertainty, calculated using a simple Monte Carlo method, as described in the text. Both quantities are shown to be systematically raised over the mean value out to approx. 4~\reff.}\label{fig:trad_plot}
\end{figure}

For each bubble, we identified the clumps within 20 R$_{\rm eff}$, in bins of 0.2 R$_{\rm eff}$. In each bin, the average temperature and linewidth was calculated, and this was repeated for each of the 1688 bubbles in the sample. The resulting values were averaged over all bubbles to identify any systematic trends. To estimate the uncertainty on the values, we performed a simple Monte Carlo (MC) analysis with 100 iterations. In each iteration, T$_{\rm kin, i}$ was replaced by a randomly chosen value from the normal distribution with $\mu =$ T$_{\rm kin,i}$ and $\sigma$ = $\Delta$T$_{\rm kin,i}$ where $\Delta$T$_{\rm kin,i}$ was calculated in \citet{Wienen2012} assuming a Gaussian error distribution on the main beam temperature and standard error propagation formulae. Similarly, for the linewidths we perform 100 MC iterations with the linewidth of each clump in each iteration drawn from the normal distribution with $\mu =$ $\Delta v_{i}$ and $\sigma$ = $\Delta(\Delta v)_{\rm i}$. The result of these calculations are shown in Figure~\ref{fig:trad_plot} for temperature and NH$_3$ (1,1) linewidth, to 4~\reff. 

The temperature of the dust clumps is raised above the average of the full sample (20.7 K) at a statistically significant level ($\geq$ 3-$\sigma$) to 4~\reff. Given that the kinetic temperature is a derived quantity the uncertainty is somewhat larger than those of the linewidths, which are measured directly from the spectral data. The increased linewidth is significant at $\geq$ 4-$\sigma$ to 12~\reff, with the increase measuring 2-10\% higher than the mean of 2.17 km s$^{-1}$, as with the temperature. This suggests that the dense material in the vicinity of expanding infrared bubbles is significantly hotter and more turbulent out to large distances beyond the bubble rims. The observed increased temperatures do not account for the increased linewidth, indicating that both the thermal and non-thermal components of the linewidth are larger near bubbles. We note that we cannot determine whether this is a consequence of internal or external factors based on these numbers alone.

\begin{figure}
    \includegraphics{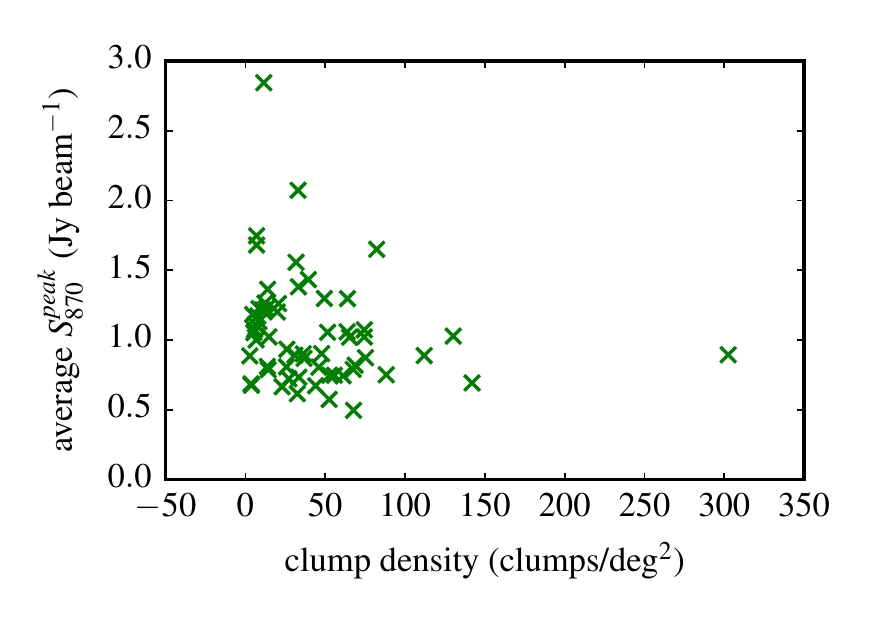}
    \caption{The peak fluxes of all ATLASGAL clumps averaged in $2 \times 2\degree$ bins in longitude and latitude, regardless of their proximity to a bubble, plotted against the clump density in the bin. No correlation is apparent, indicating that the  different fluxes seen for bubble-associated and control clumps are not a result of density of the local environment.}\label{fig:clump_phys_dens}
\end{figure}

We test whether the different physical conditions found in the clump populations are a function of their location in the Galactic Plane. Could the increased turbulence and temperature simply reflect the denser environments in which these clumps reside, e.g. within a spiral arm, regardless of the proximity to an IR bubble? In Figure~\ref{fig:clump_phys_dens} we plot the peak fluxes (related to the H$_2$ column density via Equation~\ref{eq:coldens}) of ATLASGAL clumps averaged in $2 \times 2\degree$~bins in longitude and latitude against the clump density of the bin (number of clumps/deg$^2$), without regard to the clumps' proximity to a bubble. The plot reveals no correlation between these quantities, suggesting that the difference in physical properties is not simply due to local density and the proximity to bubbles is indeed relevant.

\section{Discussion}

\subsection{Distribution of cold dust and gas near infrared bubbles}

The statistical analyses presented above show that the cold dense dust clumps are strongly clustered near infrared bubbles. Assuming that most bubbles represent expanding~\hiirs~surrounding single or small clusters of OB-type stars, this implies that quiescent and star-forming clouds are co-located with more evolved stages of star formation. Previous work has shown that ATLASGAL clumps are sufficiently massive to host the formation of high-mass stars, and that they span a range in evolutionary stages, from quiescent pre-stellar clumps to actively star-forming. Similarly, bubbles can be seen around newly formed stars over a range of ages, typically from the UCHII region to evolved diffuse~\hiir~stages (and a small number around supernovae). Seeing these objects consistently co-located throughout the Galactic Plane, in higher numbers than would be expected from random distributions, suggests that the majority of star forming regions host objects covering several Myr in age spread and that star formation in molecular clouds proceeds in a phased manner rather than in a single star formation episode. This finding is consistent with previous observational studies of single star forming complexes~\citep{Deharveng2008, Povich2009, Bik2010a, Povich2011, Wright2010}, and our analysis essentially demonstrates that this is observed statistically, throughout the inner Galactic Plane. Whilst the level of chance alignments given random distributions of these datasets on sky is not negligible (15-20\%), the observed association of 48\% of cold clumps with bubbles is clearly significant.

For the full sample of bubbles, we find a strong positional correlation with cold clumps everywhere in the bubble interior. This mirrors our findings from ~\citet{Kendrew2012}, where we compared the locations of bubbles with those of YSOs and (UC)\hiirs. The overdensity persists to approximately 2~\reff, beyond which no more clumps are seen than are expected from a random distribution. The clumps beyond 1~\reff~may represent material swept up by the bubble's expansion or pre-existing dense clumps that have stalled the expansion process. 

Clumps projected towards bubble interiors ($<$ 1 \reff) may be associated with the front and rear surfaces of the bubbles, however~\citet{Beaumont2010} found little evidence of this in an observational study of CO and HCO$^+$.  For some highly elliptical bubbles, the~\reff~as defined in Equation~\ref{eq:reff} may be a poor approximation of the bubble's shape. The fact that smaller bubbles have stronger overdensities of clumps towards their interiors than their larger counterparts could indicate that some very young stars or clusters driving bubble expansion may still be associated with their nascent massive clump, mirroring the RMS/MWP evolutionary degeneracy we found in~\citet{Kendrew2012}. Finally, clumps projected towards bubble interiors may well be line-of-sight interlopers. 

The correlation function ranked by clump column density shows that the clumps with the highest column densities ($\geq 1 \times 10^{22}$ cm$^{-2}$) are significantly overdense towards bubble interiors. Along the bubble rims these clumps are only marginally overdense compared with the full clump sample. This seems to suggest that a significant fraction of these high N(H$_2$) clumps are located within the bubble interior. Interestingly recent magnetohydrodynamic simulations of~\hiir~expansion into turbulent clouds by~\citet{Geen2015} show that the most massive of clumps are able to resist the expansion of an~\hiir~and remain inside the bubble radius as cometary clouds; this may explain our observed correlation.

When performing the analysis for sub-samples of bubbles of increasing (angular) size, the correlation function decreases at small angular separations (i.e. towards the bubble interior), but remains high around the bubble radius. Assuming that large bubbles are more evolved than their smaller counterparts at any given distance, we interpret this as the statistical signature of dense material being cleared out of the bubbles' interior, and swept-up material piling up along the rim, as they expand and evolve. This is consistent with our results in ~\citet{Kendrew2012}, and with the findings of numerous studies of individual star forming regions~\citep{Zavagno2010a}.

A similar study by~\citet{Hou2014} correlating the distribution of molecular gas, as traced by $^{13}$CO find a similar peak in both azimuthally averaged $^{13}$CO flux density and clump number counts near the rims of MWP bubbles. They attempt to correct for projection effects and estimate the fraction of bubble-associated clumps to be $\sim$20\%. Given the different datasets, gas tracers and methodology used, it is hard to compare our results with this study in a consistent manner, however our conclusions are broadly consistent.

The analytical framework of~\citet{whitworth94, Elmegreen1977} predicts that the material swept up and compressed by the expanding shell will become gravitationally unstable, fragment, and collapse to form new stars; the overdensity of YSOs found in ~\citet{Kendrew2012} and by~\citet{thompson12},~\citet{Urquhart2014} and~\citet{deharveng10} amongst others supports this theory. This shell fragmentation could possibly explain the increased number count of clumps around the bubble rims. However when we compare the sizes of clumps near bubbles compared with their counterparts in the field, we see this is not the case: both samples have indistinguishable size distributions. We can therefore conclude that there is simply more cold dense gas and dust near bubbles.

\subsection{Star formation near infrared bubbles}

Numerous studies, including our own, have found star formation to be enhanced in the vicinity of IR bubbles. While we cannot determine whether ATLASGAL clumps are actively forming stars from the 870~\micron~emission alone, we can make inferences from the clumps' physical properties and from complementary studies in the literature. By performing an initial cross-matching with the mid-infrared (mid-IR) point source catalogues from MSX (21.3~\micron) and WISE (22~\micron), \citet{Csengeri2014} suggest that at least 33\% of clumps have embedded protostellar objects. They find that the fraction of star forming clumps increases with flux density, reaching a saturation level at $S_{870}^{peak} > 5$ Jy beam$^{-1}$, where 75\% of all clumps are associated with mid-IR point sources. In our bubble-associated clump sample, 129 clumps have peak fluxes above this value (97th percentile); the control sample contains only 36 clumps with peak fluxes $> 5$ Jy beam$^{-1}$ (99th percentile). This suggests that the clumps near bubbles are more likely to be forming stars than their counterparts in the field and/or that the stars that form are more massive than those in field clumps.

Similarly, in a census of star formation activity (as traced by mid-infrared emission) associated with massive cold clumps from the 1.1-mm Bolocam Galactic Plane Survey (BGPS;~\citet{Aguirre2011}),~\citet{Dunham2011} find higher millimeter fluxes and mean H$_2$ column densities to be associated with an increased probability of star formation activity. The BGPS clump sample traces similar sources to those from ATLASGAL, albeit at lower spatial resolution ($\sim$30\arcsec) and over a smaller survey area. Our findings presented in this paper are qualitatively consistent with those of~\citet{Dunham2011}; their work backs the notion that the ATLASGAL clumps found near IR bubbles are more likely to be forming stars than those in the field. This is in agreement also with ~\citet{Kendrew2012}.

We did not attempt to divide our clump sample into quiescent and star-forming sub-samples for this study. The work of~\citet{Schuller2009, Tackenberg2012, Dunham2011} has shown the difficulty with a conclusive identification of star formation activity at all evolutionary stages. Far-IR data are typically required to unequivocally constrain the spectral energy distributions of young stellar objects; the forthcoming source catalogue from the far-infrared HiGAL survey~\citep{Molinari2010} will help with the study of star formation in dense cold clumps.

\subsection{The effect of bubble feedback}

Our work has shown that cold dust clumps in the vicinity of IR bubbles are warmer, more turbulent and with higher H$_2$ column densities than those in the field, on spatial scales beyond the immediate shell swept up by the bubble expansion. These are two possible manifestations of massive stellar feedback: (i) the large-scale dispersal of material surrounding the central star (the bubble expansion), and subsequent fragmentation and collapse of the shell; and (ii) the injection of turbulent energy into the surrounding ISM. The relative importance of these two mechanisms is likely to depend on the density and structure of the surrounding cloud. The increase is of the order of 10\%. 

The increased probability of bubble-associated clumps to be forming massive stars, shown in the previous section, may be interpreted as evidence of feedback-driven star formation. However, our analysis cannot conclusively establish causality. The larger linewidths and higher temperatures seen near bubbles \emph{may} be evidence of increased turbulence or heating injected by feedback from the bubbles' driving sources, it may be due to internal heating by nascent star formation, or indeed or a combination of several effects. 

Interestingly we show that the temperature and NH$_3$ linewidths are raised over the global average to large distances beyond the bubble rims ($\leq$ 4-5~\reff). This hints at the large spatial scales over which massive stellar feedback, in the form of photoionization, stellar winds and radiation pressure, can act to affect physical conditions in the cloud. This phenomenon has been noted in both observational studies, e.g.~\citet{Murray2010}'s work correlating star formation observables between the GLIMPSE and WMAP surveys, or~\citet{Lopez2014}' work on~\hiirs~in the Magellanic Clouds; as well as in simulations~\citep{Dale2012, Dale2013, Geen2015}. Our results allow us to visualise and begin to quantify this on Galactic scales.

\section{Conclusion}

We have presented here a detailed statistical study of the distribution and physical properties of massive cold dust clumps in the inner Galactic Plane, as detected in the ATLASGAL survey, in correlation with interstellar bubbles from the Milky Way Project, marking the likely locations of newly formed massive young stars and clusters. We can summarise our conclusions as follows:

\begin{itemize}
    \item In a simple proximity analysis, we find that almost half (48 $\pm$ 2\%) of cold dust clumps in or near the inner Galactic Plane are located in the vicinity of a MWP bubble, and a quarter (26 $\pm$ 2\%) near a bubble rim. Both these values lie significantly above the rate of chance alignments.
    \item The angular correlation analysis between the bubble and clump samples show an overdensity of dust clumps towards bubbles, and a marked overdensity of clumps near bubble rims for increasing bubble size. This observation cannot be explained by increased clump fragmentation near bubble rims. We interpret this as the statistical evidence of the large-scale reordering of ISM material as a consequence of the expansion of bubbles around regions of massive star formation. There is evidence that the clumps with the highest column densities are able to resist being swept up by the expanding bubbles, remaining in the bubble interior.
    \item The ATLASGAL sources represent massive clumps of cold dust and gas, potentially hosting massive star formation at a wide range of evolutionary stages from quiescent to relatively evolved~\citep{Csengeri2014}. This, in combination with the above findings, leads us to conclude that the majority of star forming regions in the inner Galactic Plane harbour a wide range of evolutionary stages from pre-collapse clumps to evolved~\hiirs, and that star formation proceeds in a phased manner throughout the inner Galactic Plane. 
    \item We find differences in physical properties between bubble-associated cold clumps and their counterparts in the field as calculated from NH$_3$ spectroscopy of a subset of ATLASGAL clumps. Clumps near bubbles are significantly hotter and more turbulent and display higher H$_2$ column densities than the field clumps. For the column density measurements we do not find the difference to reflect the local density, suggesting that the proximity to a bubble is indeed relevant. When compared with complementary studies of cold dust clumps in the literature~\citep{Dunham2011}, we conclude that the bubble-associated clumps are more likely to be forming stars than those in the field, and/or that the stars that form are more massive. This is consistent with our results from ~\citet{Kendrew2012}, which found an overdensity of MYSOs in the vicinity of MWP bubbles.
    \item Studying the evolution of clumps' physical properties with radial distance from the bubbles, we observe that temperatures and NH$_3$ (1,1) linewidths are raised above the sample average to $<$ 4~\reff, i.e. beyond the range where the overdensity in pure number counts is observed. This suggests that the possible feedback effect from the young stars' ionizing radiation and the bubble expansion penetrates the ISM material to substantial distances, driving turbulence and heating in the gas, and perhaps ultimately contributing to the destruction of the cloud.
    \item Whilst our results do not prove or disprove the existence of triggered star formation near bubbles, the evidence presented supports the findings of ~\citet{Kendrew2012}, that massive star formation is significantly enhanced in the vicinity of IR bubbles, with the enhancement increasing near the largest (i.e. most evolved) bubbles.
\end{itemize}

\section{Code}\label{sec:code}
                                    
The main body of code to perform the correlation and auto-correlation analyses presented in this paper was written in Python version 2.6/2.7 and makes use of the Astropy package~\citep{astropy}. It is freely available for download as a public Github repository\footnote{https://github.com/skendrew}. We invite and encourage other authors to download the code, reuse or improve it for reproduction of these results or for similar analyses. 

\acknowledgements{

This publication has been made possible by the participation of more than 35,000 volunteers on the Milky Way Project. Their contributions are acknowledged individually at http://www.milkywayproject.org/authors. Images in the Milky Way Project are based on observations made with the Spitzer Space Telescope, which is operated by the Jet Propulsion Laboratory, California Institute of Technology under a contract with NASA. We also use data acquired with the Atacama Pathfinder EXperiment (APEX). APEX is a collaboration between the Max Planck Institute for Radioastronomy, the European Southern Observatory, and the Onsala Space Observatory. 

SK made extensive use of NASA's Astrophysics Data System Bibliographic Services for this work, and is grateful for helpful discussions with Mark Thompson, James Urquhart, Janet Drew, Jim Dale, Grace Wolf-Chase and Eve Ostriker. TCs acknowledges financial support from the ERC Advanced Grant GLOSTAR under contract no. 247078.

\facility{
{\it Facilities:} \facility{Spitzer (IRAC, MIPS)}, \facility{APEX}
}



\bibliographystyle{apj}
\bibliography{mwp-triggering}

\end{document}